\documentclass[10pt,conference]{IEEEtran}
\usepackage{graphicx}
\usepackage{amsmath}

\begin{document}

\title{Simulation and Implementation of Decoy State Quantum Key Distribution over 60km Telecom Fiber}

\author{\authorblockN{Yi Zhao, Bing Qi, Xiongfeng Ma, Hoi-Kwong Lo, Li Qian}
\authorblockA{Center for Quantum Information and Quantum Control\\
Department of Physics and Department of Electrical \& Computer
Engineering\\
 University of Toronto, Toronto, Ontario, M5S 3G4, CANADA\\
 Email: yzhao@physics.utoronto.ca, hklo@comm.utoronto.ca} }
% avoiding spaces at the end of the author lines is not a problem with
% conference papers because we don't use \thanks or \IEEEmembership
%Homer Simpson\authorrefmark{2},
%Montgomery Scott\authorrefmark{3} and
%Eldon Tyrell\authorrefmark{4}}
%Telephone: (800) 555--1212, Fax: (888) 555--1212}
%\authorblockA{\authorrefmark{4}Tyrell Inc., 123 Replicant Street, Los Angeles, California 90210--4321}}

\maketitle

\begin{abstract}
Decoy state quantum key distribution (QKD) has been proposed as a
novel approach to improve dramatically both the security and the
performance of practical QKD set-ups. Recently, many theoretical
efforts have been made on this topic and have theoretically
predicted the high performance of decoy method. However, the gap
between theory and experiment remains open. In this paper, we report
the first experiments on decoy state QKD, thus bridging the gap. Two
protocols of decoy state QKD are implemented: one-decoy protocol
over 15km of standard telecom fiber, and weak+vacuum protocol over
60km of standard telecom fiber. We implemented the decoy state
method on a modified commercial QKD system. The modification we made
is simply adding commercial acousto-optic modulator (AOM) on the QKD
system. The AOM is used to modulate the intensity of each signal
individually, thus implementing the decoy state method. As an
important part of implementation, numerical simulation of our set-up
is also performed. The simulation shows that standard security
proofs give a zero key generation rate at the distance we perform
decoy state QKD (both 15km and 60km). Therefore decoy state QKD is
necessary for long distance secure communication. Our implementation
shows explicitly the power and feasibility of decoy method, and
brings it to our real-life.
\end{abstract}

\section{Introduction}
Quantum key distribution (QKD)\cite{BB84,ekert1991} was proposed as
a method of sharing a key between two parties (normally denoted by
the sender Alice and receiver Bob) securely. It would be impossible
for an eavesdropper to attack the QKD system without being detected.
Assuming a perfect single photon source is utilized, people have
proven the security of QKD based on fundamental laws of quantum
physics \cite{securityproof, moresecurityproof}.

Unfortunately, in view of implementation, ``perfect'' devices are
always very hard to build. Therefore most up-to-date QKD systems
substitute the desired perfect single photon sources by heavily
attenuated coherent laser sources. QKD can be performed with these
laser sources over more than 120km of telecom fibers \cite{GYS,Guo}.

However, this substitution raises some severe security concern. The
output of coherent laser source obeys Poisson distribution. Thus the
occasional production of multi-photon signals is inevitable no
matter how heavily people attenuate the laser. Recall that the
security of BB84 protocol \cite{BB84} is guaranteed by quantum
no-cloning theorem, the production of multi-photon signals is fatal
for the security: the eavesdropper (normally denoted by Eve) can
simply keep an identical copy of what Bob possesses by blocking all
single-photon signals and splitting all multi-photon signals. Most
up-to-date QKD experiments have not taken this photon-number
splitting (PNS) attack into account, and thus are, in principle,
insecure.

Is it possible for people to develop some special measure to make
QKD secure even with practical systems? The answer is yes. From
physical intuition, if Alice sends out a single photon signal, and
Bob luckily receives it, this bit (normally defined as in ``single
photon state'') should be secure, because Eve cannot split or clone
it. Based on this intuition, rigorous security analysis on practical
QKD system is proposed by \cite{ilm} and
Gottesman-Lo-L\"{u}tkenhaus-Preskill (GLLP)\cite{GLLP}, which is
based on the entanglement distillation approach to security proofs.

The main idea of GLLP's work is not to find \emph{which} signals are
secure (i.e., single-photon signals), because it would be beyond
current technology. Instead, GLLP shows that the \emph{ratio} of
secure signals can be estimated from some experimental parameters,
and secure key bits can then be extracted from the raw key based on
this ratio through data post-processing.

With the GLLP method the secure key generation rate, which is
defined as the ratio of the length of the secure key to the total
number of signals sent by Alice, is given by  \cite{GLLP}
\begin{equation}\label{R}
R\geq q\{-Q_\mu f(E_\mu)H_2(E_\mu)+Q_1[1-H_2(e_1)]\},
\end{equation}
where $q$ depends on the protocol; the subscript $\mu$ is the
average photon number per signal in signal states; $Q_\mu$ and
$E_\mu$ are the gain and the quantum bit error rate (QBER)  of
signal states, respectively; $Q_1$ and $e_1$ are the gain and the
error rate of the single photon state in signal states,
respectively; $f(x)$ is the bi-directional error correction rate
 \cite{brassard1994}; and $H_2(x)$ is the binary entropy function:
$H_2(x)=-x\log_2(x)-(1-x)\log_2(1-x)$. $Q_\mu$ and $E_\mu$ can both
be measured directly from experiments, while $Q_1$ and $e_1$ have to
be estimated (because Alice and Bob could not measure the photon
number of each pulse with current technology).

GLLP \cite{GLLP}  has also given a method to estimate the lower
bound of $Q_1$ and the upper bound of $e_1$, thus giving out the
lower bound of the key rate $R$. However, with coherent laser
sources, these bounds are not tight. It follows that the security of
practical QKD set-ups can be guaranteed only at very short distance
and very low key generation rate \cite{GLLP,decoy}.

A key question is thus raised: how can one extend both the maximum
secure distance and key generation rate of QKD? The most intuitive
choice would be to use a (nearly) perfect single photon source.
Despite much experimental effort \cite{singlephoton}, reliable
near-perfect single photon sources are far from practical.

Another solution to increase the maximum secure distance and key
generation rate is to employ decoy states, using extra states of
different average photon number to detect photon-number dependent
attenuation. The decoy method was first discovered by Hwang
\cite{hwang2003}. The first rigorous security proof of decoy state
QKD was presented by us \cite{decoy}. It is shown that the decoy
state method can be combined with standard GLLP result to achieve
dramatically higher key generation rates and longer distances
\cite{decoy}. Moreover, practical protocols with vacua and weak
coherent states as decoys were proposed \cite{decoy}. Subsequently,
we have analyzed the security of practical protocols
\cite{practical}. Decoy method has attracted great recent interests
 \cite{wang}.

The basic idea of decoy state QKD is as follows: Alice introduces
some some ``decoy'' states with average photon numbers $\nu_i$
besides the signal state with average photon number $\mu$
($\neq\nu_i$). Each pulse sent by Alice is assigned to a state
(signal state or one of the decoy states) randomly. Alice announces
the state of each pulse after Bob's acknowledgement of receipt of
signals. The statistical characteristics (i.e., gain and QBER) of
each state can then be analyzed separately. Note that the average
photon number of certain state is only by statistical meaning, while
Eve's knowledge is limited to the actual photon number in each
pulse, therefore Eve has no clue about the state of each pulse.
Eve's attack will modify the statistical characteristics (gain or
QBER) of decoy states and/or signal state and will be caught. The
decoy states are used only for catching an eavesdropper, but not for
key generation. It has been shown \cite{decoy,practical,wang} that,
in theory, decoy state QKD can substantially enhance security and
performance of QKD.

The power and feasibility of decoy method can be shown only by
implementing it. To implement decoy state QKD, it is intuitive to
utilize variable optical attenuators (VOAs) to modulate the
intensity of each signal to that of its state. Actually, this is
exactly the way we used.

\section{Implementations of Decoy State Protocols}

In \cite{decoy,practical}, we have proposed several protocols on
decoy state QKD. The most important two protocols are the one-decoy
protocol (the simplest protocol) and the weak+vacuum protocol (the
optimal protocol). We have implemented both of them, over 15km (the
one-decoy protocol) and 60km (the weak+vacuum protocol) standard
telecom fibers.

\subsection{Implementation of one-decoy
protocol}\label{se:one-decoy}

\begin{figure}[!t]\center
\resizebox{7.5cm}{!}{\includegraphics{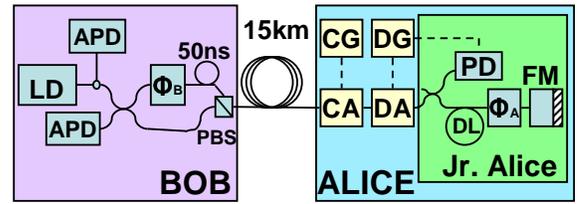}}
\caption{Schematic of the set-up in one-decoy protocol experiment.
Inside Bob/Jr. Alice: components in Bob/Alice's package of id
Quantique QKD system. Our modifications: CA: Compensating AOM; CG:
Compensating Generator; DA: Decoy AOM; DG: Decoy Generator. Original
QKD system: LD: laser diode; APD: avalanche photon diode; $\Phi_i$:
phase modulator; PBS: polarization beam splitter; PD: classical
photo detector; DL: Delay line; FM: faraday mirror. Solid line:
SMF28 single mode optical fiber; dashed line: electric
cable.}\label{schematic_one}
\end{figure}

In one-decoy protocol, people need only \emph{one} decoy state with
average photon number per signal $\nu<\mu$. Alice could decide the
values of $\mu$ and $\nu$, and the ratio of number of pulses used as
decoy state to that of total pulses, then randomly assign the state
to each signal by attenuating the intensity of each signal to either
$\mu$ or $\nu$.

We implemented the one-decoy protocol by adding acousto-optical
modulators (AOMs, including CA, DA in Fig. \ref{schematic_one}) to a
commercial ``Plug \& Play'' QKD system manufactured by id Quantique
(Jr. Alice and Bob in Fig. \ref{schematic_one}). We choose AOM to
modulate the signals because we need this amplitude
modulation to be polarization insensitive.%This QKD system is
%bi-directional in the sense that Bob sends out a chain of strong
%signals to Alice, who attenuates each signal to single photon level
%and modulates (i.e., encodes quantum information on) its phase
%before sending it back to Bob, who performs the measurement (i.e.,
%decoding of quantum information), after which a new chain of strong
%signals will be sent to Alice.

This QKD system is based on a 1550nm laser source with pulse
repetition rate of 5MHz. Its intrinsic parameters, including dark
count rate $Y_0$, detector error rate $e_{detector}$, and Bob's
quantum efficiency $\eta_{Bob}$ are listed on Table \ref{intrisic}.

\begin{table}[!b]\center
\caption{Some intrinsic parameters of the QKD system. These
parameters are different for the two implementations because the
single photon detectors of the QKD system were adjusted by the
manufacturer between the two experiments.}\label{intrisic}
\begin{tabular}{l c c c}
\hline
Implementation & $Y_0$ & $e_{detector}$ & $\eta_{Bob}$\\
\hline One-Decoy & $2.11\times10^{-5}$ &
$8.27\times10^{-3}$ & $2.27\times10^{-2}$\\
\hline Weak+Vacuum & $6.14\times10^{-5}$ & $1.38\times10^{-2}$ &
$5.82\times10^{-2}$\\
\hline

\end{tabular}

\end{table}

Before experiment, we perform a numerical simulation (discussed in
detail in Section \ref{se:simulation}) with parameters of our set-up
as in Table \ref{intrisic} and optimally set $\mu$ and $\nu$ to 0.80
and 0.120 photons respectively. The actual distribution of the
states is produced by an id Quantique Quantum Random Number
Generator. Around 10\% of the signals are assigned as decoy states
as suggested by numerical simulation. This random pattern is
generated and loaded to the Decoy Generator (DG in Fig.
\ref{schematic_one}) before the experiment.

Here we describe the flow of the experiment. First, Bob generates a
chain of strong laser pulses by the laser diode (LD in Fig.
\ref{schematic_one}) and sends them to Alice through the 15km fiber.
Second, the pulses propagate through the AOMs (CA and DA in Fig.
\ref{schematic_one}, the function of CA as well as CG will be
discussed in the next paragraph), whose transmittances are set
maximum at this period. Third, each pulse is splitted by a coupler
and part of it will be detected by a classical photo detector (PD in
Fig. \ref{schematic_one}), which generates synchronizing signal to
trigger the Decoy Generator (DG in Fig. \ref{schematic_one}).
Fourth, the generator holds for certain time period, during which
the pulses are reflected by the faraday mirror (FM in Fig.
\ref{schematic_one}) and quantum information is encoded by the phase
modulator ($\Phi_A$ in Fig. \ref{schematic_one}). Here comes the key
point: fifth, the Decoy Generator (DG in Fig. \ref{schematic_one})
will drive the Decoy AOM (DA in Fig. \ref{schematic_one}) to
modulate each pulse to the intensity (either 0.80 or 0.120) of the
state it is assigned to exactly when the pulse propagates through
the AOM. Sixth, the pulses (now in single photon level) return to
Bob through the 15km fiber again. Seventh, Bob decodes the quantum
information by modulating the phases of the pulses by the phase
modulator ($\Phi_B$ in Fig. \ref{schematic_one}) and see which
single photon detector (APD in Fig. \ref{schematic_one}) fires.

The use of the Decoy AOM (DA in Fig. \ref{schematic_one}) shifts the
frequency of the laser pulses, thus shifts the relative phase
between pulses significantly. To compensate this phase shift,
another AOM, the ``Compensating AOM'' (CA in FIG.
\ref{schematic_one}) is employed to make the total phase shift
multiples of $2\pi$. This AOM is driven by the second function
generator, ``Compensating Generator'' (CG in FIG.
\ref{schematic_one}). Its transmittance is set constant throughout
the experiment.

Here we emphasize that the holding time of the Decoy Generator (DG
in Fig. \ref{schematic_one}) after being triggered by the photo
detector (PD in Fig. \ref{schematic_one}) must be very precise,
because same modulation must be applied to the two pulses of the
same signal to keep visibility high. In our experiment, the
precision of this holding time is 10ns.

After the transmission of all the signals, Alice broadcasted to Bob
the distribution of decoy states as well as basis information. Bob
then announced which signals he had actually received in correct
basis. We assume Alice and Bob announced the measurement outcomes of
all decoy states as well as a subset of the signal states. From
those experimental data, Alice and Bob then determined $Q_\mu$,
$Q_\nu$ $E_\mu$, and $E_\nu$, whose values are now listed in
Table~\ref{expresult_one}. Note that our experiment is based on BB84
\cite{BB84} protocol, thus $q=N_\mu^S/N$, where $N_\mu^S$ is the
number of pulses used as signal state when Alice and Bob chose the
same basis, and $N=105$Mbit is the total number of pulses sent by
Alice in this experiment.

Now we have to analyze the experimental result and estimate the
lower bound of key generation rate $R$. This can be done by simply
inputting the results in Table \ref{expresult_one} to the following
equations \cite{practical}:

\begin{equation}\label{eq:Q1e1_onedecoy}
\begin{aligned}
Q_1 &\ge Q_1^L = \frac{\mu^2e^{-\mu}}{\mu\nu-\nu^2}(Q_\nu^L
e^{\nu}-Q_\mu e^\mu\frac{\nu^2}{\mu^2}-E_\mu Q_\mu e^\mu
\frac{\mu^2-\nu^2}{e_0\mu^2})\\
e_1 &\le e_1^U = \frac{E_\mu Q_\mu}{Q_1^{L}},
\end{aligned}
\end{equation}
in which
\begin{equation}\label{eq:QnuL}
%\begin{aligned}
Q_\nu^L = Q_\nu(1-\frac{u_\alpha}{\sqrt{N_\nu Q_\nu}}),
%\end{aligned}
\end{equation}
where $N_\nu$ is the number of pulses used as decoy states, and
$e_0$ (=1/2) is the error rate for the vacuum signal and therefore
the lower bound of key generation rate is
\begin{equation}\label{eq:RL}
R \ge R^L=q\{-Q_\mu f(E_\mu)H_2(E_\mu)+Q_1^L[1-H_2(e_1^U)]\}
\end{equation}

In our analysis of experimental data, we estimated $e_1$ and $Q_1$
very conservatively as within 10 standard deviations (i.e.,
$u_\alpha$=10), which promises a confidence interval for statistical
fluctuations of $1-1.5\times10^{-23}$.

Even with our very conservative estimation of $e_1$ and $Q_1$, we
got a lower bound for the key generation rate $R^L=3.6\times10^{-4}$
per pulse, which means a final key length of about
$L=NR\simeq38$kbit. We also calculated
$R_{perfect}=1.418\times10^{-3}$, the theoretical limit from the
case of infinite data size and infinite decoy states protocol, by
using Eq. (1). We remark that our lower bound $R^L$ is indeed good
because it is roughly $1/4$ of $R_{perfect}$.

%Based on the method described in  \cite{GLLP,decoy,practical}, we
%carefully performed numerical simulations based on data in Table
%\ref{intrisic}. We found that without decoy method, no matter what
%value of $\mu$ we choose or how large the data size we use, the key
%generation rate, $R$,  will hit zero at only 9.5km. In other words,
%at 15km, not even a single bit could be shared between Alice and Bob
%with guaranteed security. In contrast, our numerical simulations
%show that, with decoy states, our QKD set-up can be made secure over
%50km, which is substantially larger than the secure distance (9.5km)
%without decoy states.

\begin{table}[!b]\center
\caption{Experimental results in one-decoy protocol. As required by
GLLP \cite{GLLP}, bit values for double detections are assigned
randomly by the quantum random number
generator.}\label{expresult_one}
\begin{tabular}{c c |c c| c c}
\hline

Para. & Value & Para. & Value & Para. & Value\\

\hline $Q_{\mu}$ &$8.757\times10^{-3}$ &$E_{\mu}$
&$9.536\times10^{-3}$ &$q$&0.4478\\
 $Q_{\nu}$ &$1.360\times10^{-3}$ &$E_\nu$&$2.689\times10^{-2}$
&$f(E_\mu)$ \cite{brassard1994}& $\leq$1.22\\
\hline

\end{tabular}

\end{table}

\subsection{Implementation of weak+vacuum protocol}
\begin{figure}[!t]\center
\resizebox{7.5cm}{!}{\includegraphics{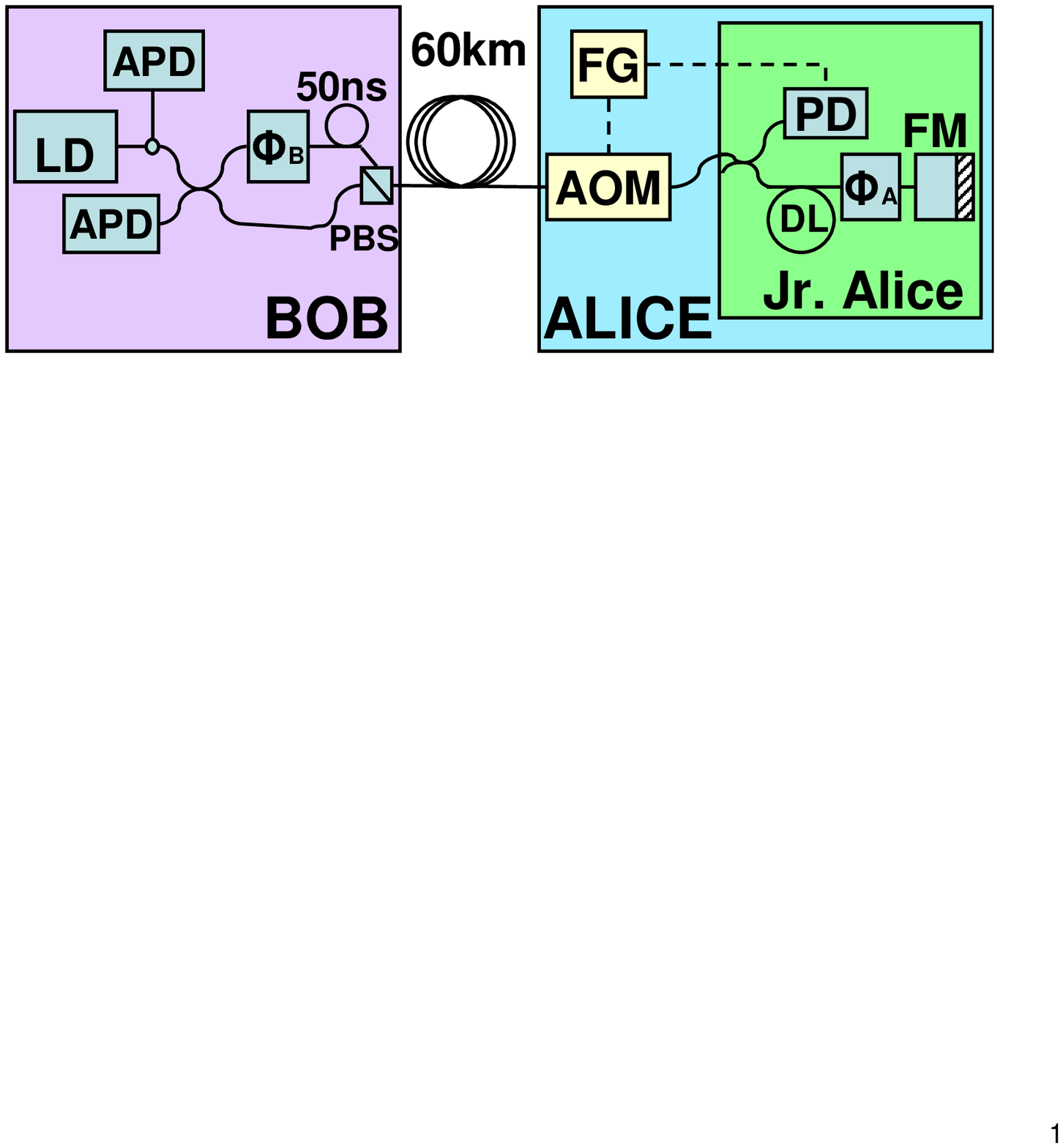}}
\caption{Schematic of the set-up in weak+vacuum protocol experiment.
Inside Bob/Jr. Alice: components in Bob/Alice's package of id
Quantique QKD system. Our modifications: AOM: Decoy AOM; FG:
Functional Generator. Original QKD system: LD: laser diode; APD:
avalanche photon diode; $\Phi_i$: phase modulator; PBS: polarization
beam splitter; PD: classical photo detector; DL: Delay line; FM:
faraday mirror. Solid line: SMF28 single mode optical fiber; dashed
line: electric cable.}\label{schematic_wv}
\end{figure}

Weak+Vacuum protocol is similar to one-decoy protocol except that it
has one more decoy state: the vacuum state, which has zero
intensity. The vacuum state is to detect the background count rate.
We hereby use the same notation for intensities as Subsection
\ref{se:one-decoy}: $\mu$ for signal state and $\nu<\mu$ for weak
decoy state.

Weak+Vacuum protocol is theoretically predicted to have higher
performance than one-decoy protocol and is optimal protocol in
asymptotic case \cite{decoy,practical}. Our numerical simulation
(detailed discussion in Section \ref{se:simulation}) shows that for
our set-up (as in Table \ref{intrisic}), with data size of 105Mbit,
the maximum secure distance for one-decoy protocol is 59km, while
that of weak+vacuum protocol is 68km, as shown in Fig.
\ref{fig:simulation-wv}. We chose 60km telecom fiber to perform
weak+vacuum protocol.

\begin{table}[!b]\center
\caption{The experimental results of weak+vacuum
protocol.}\label{tab:expresult_wv}
\begin{tabular}{c c | c c}
\hline Para. & Value & Para. & Value\\
\hline $Q_\mu$ & $1.81\times10^{-3}$ & $E_\mu$ & $3.05\times10^{-2}$\\
\hline $Q_\nu$ & $5.47\times10^{-4}$ & $E_\nu$ & $7.78\times10^{-2}$\\
\hline $Y_0$ & $6.02\times10^{-5}$ & $e_0$ & $0.51$\\
\hline $q$ & $0.319$ & $f(E_\mu)$\cite{brassard1994} & $\le1.22$\\
\hline
\end{tabular}

\end{table}

The implementation of weak+vacuum protocol requires amplitude
modulation of three levels: $\mu$, $\nu$ and 0. Note that it would
be quite hard for high-speed amplitude modulators to prepare the
real ``vacuum'' state due to finite distinction ratio. However, if
the gain of the ``vacuum'' state is very close (like within a few
standard deviations) to the dark count rate, it would be a good
approximation.

Our set-up to implement weak+vacuum protocol (Fig.
\ref{schematic_wv}) is very similar to that of one-decoy protocol
(Fig. \ref{schematic_one}) except for the absence of the
``compensating'' parts (CA \& CG in Fig. \ref{schematic_one}). This
is because the frequency of the AOM (AOM in Fig. \ref{schematic_wv})
has been precisely adjusted to the value that the phase shift caused
by it is exactly multiples of $2\pi$. In other words, this AOM is
self-compensated for our set-up.

We performed numerical simulation (as discussed in details in
Section \ref{se:simulation}) to find out the optimal parameters.
According to simulation results, we choose the intensities as
$\mu=0.55$, $\nu=0.152$. Numbers of pulses used as signal state,
weak decoy state and vacuum state are $N_\mu=0.635N$,
$N_\nu=0.203N$, and $N_0=0.162N$, respectively, where $N=105$Mbit is
the total data size we used.

The experimental results are shown in Table \ref{tab:expresult_wv}.
Note that the gain of vacuum state ($Y_0$ in Table
\ref{tab:expresult_wv}) is indeed very close to the dark count rate
($Y_0$ in Table \ref{intrisic}, third row), therefore the vacuum
state in our experiment is quite ``vacuum''. We could estimate the
lower bound of $Q_1$ and upper bound of $e_1$ by plugging these
experimental results into the following equations \cite{practical}:

\begin{equation}\label{eq:Q1e1}
\begin{aligned}
Q_1 &\ge Q_1^L = \frac{\mu^2e^{-\mu}}{\mu\nu-\nu^2}(Q_\nu^L
e^{\nu}-Q_\mu e^\mu\frac{\nu^2}{\mu^2}-Y_0^U
\frac{\mu^2-\nu^2}{\mu^2}),\\
e_1 &\le e_1^U = \frac{E_\mu Q_\mu-e_0Y_0^Le^{-\mu}}{Q_1^{L}},
\end{aligned}
\end{equation}
in which
\begin{equation}\label{eq:Y0}
\begin{aligned}
Y_0^L &= Y_0(1-\frac{u_\alpha}{\sqrt{N_0 Y_0}}),\\
Y_0^U &= Y_0(1+\frac{u_\alpha}{\sqrt{N_0 Y_0}}),
\end{aligned}
\end{equation}
and $Q_\nu^L$ takes the value as in Eq. \eqref{eq:QnuL}. Again, we
estimate $Q_1$ and $e_1$ very conservatively by setting
$u_\alpha=10$, which promises a confidence interval for statistical
fluctuations of $1-1.5\times10^{-23}$.

A lower bound of the key generation rate $R^L=8.45\times10^{-5}$ per
pulse is found by plugging the results of Eqs. \eqref{eq:Q1e1} into
Eq. \eqref{eq:RL}, which means a final key length of about
$L=NR\simeq9$kbit. Note that, one-decoy protocol cannot give out a
positive key rate at 60km as suggested by numerical simulation.
Therefore, weak+vacuum protocol is on demand at this distance.%We
%also confirm the numerical simulation result by plugging $Q_\mu$,
%$E_\mu$, $Q_\nu$ and $q$ from Table \ref{tab:expresult_wv} into Eqs.
%\eqref{eq:Q1e1_onedecoy}\eqref{eq:QnuL}\eqref{eq:RL} and found
%indeed that no positive key rate could be found.

\section{Numerical Simulation}\label{se:simulation}

\begin{figure}[!t]\center
  \includegraphics[width=8cm]{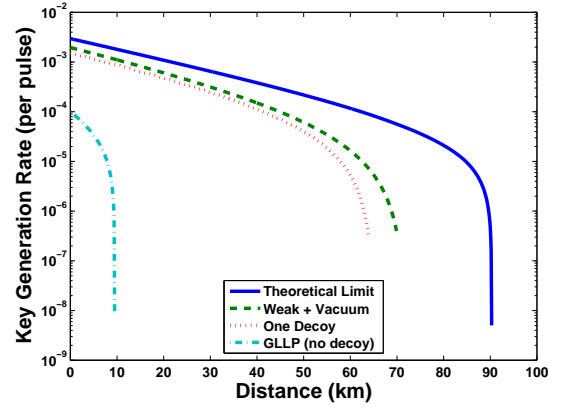}
  \caption{Simulation result of the set-up on which we implemented the one-decoy
  protocol. Intrinsic parameters for this set-up is shown in the second row of Table \ref{intrisic}.
  Solid line: the theoretical limit of key generation rate. Its maximum transmission
  distance is about 90km. Dashed line: the performance of weak+vacuum protocol. Its maximum distance is about 70km.
  Dotted line: the performance of one-decoy protocol. Its maximum distance is about 64km. Dashed and dotted line:
  the performance without decoy method. Its maximum distance is only 9.5km.}\label{fig:simulation-one}
\end{figure}

Numerical simulation is crucial for setting optimal experimental
parameters and choosing the distance to perform certain decoy method
protocol. Here we explain the principle of our simulation, and show
some results.

The principle of numerical simulation is that for certain QKD
set-up, if the intensities and percentages of signal state and decoy
states are chosen, we could simulate the experimental results (gains
and QBERs of all states). For example, suppose we have a QKD set-up
with transmittance $\eta$, detector error rate $e_{detector}$ and
dark count rate $Y_0$, if the output intensity is set to be $\mu$
photons per signal, the gain and QBER of this state is expected to
be \cite{lukenhaus2000}
\begin{equation}\label{eq:QmuEmu}
\begin{aligned}
Q_\mu &= Y_0+1-e^{-\eta\mu},\\
E_\mu &= \frac{1}{Q_\mu}(e_0Y_0+e_{detector}(1-e^{-\eta\mu})),
\end{aligned}
\end{equation}
respectively. With these simulated experimental outcome, we could
estimate the lower bound of the key generation rate.

In experiment, it is natural to choose the intensities and
percentages of signal state and decoy states which could give out
the maximum key generation rate. This search for optimal parameters
can be done by numerical simulation and exhaustive search. For
example, we could try the values of $\mu$ and $\nu_i$, the
intensities of signal state and decoy states, in the range of
$[0,1]$ with a step increase of 0.001. Similar strategy can be
applied on the percentage of each state. With certain combination of
intensities and percentages, the gains and QBERs of different states
could be simulated by Eqs. \eqref{eq:QmuEmu}, and the key generation
rate can be estimated by the chosen protocol, like Eqs.
\eqref{eq:Q1e1_onedecoy}\eqref{eq:QnuL}\eqref{eq:RL} for one-decoy
protocol and Eqs.
\eqref{eq:QnuL}\eqref{eq:RL}\eqref{eq:Q1e1}\eqref{eq:Y0} for
weak+vacuum protocol. We can therefore find out the optimal
combination that
can give maximum key generation rate. %Note that some optimization on
%algorithm is necessary to achieve the optimal combination with high
%precision in reasonable time (like in a few seconds).

Numerical simulation can also give the maximum secure distance for
certain decoy protocol and QKD set-up. The transmittance of the
system is a simple function of distance \cite{lukenhaus2000}
$\eta=\eta_{Bob}e^{-\alpha l}$, where $\alpha$(=0.21dB/km in our
set-up) is the loss coefficient. For a QKD set-up with known
$\eta_{Bob}$, $\alpha$, $e_{detector}$, and $Y_0$, we could find out
the maximum key generation rate of some protocol at distance $l$.
The shortest distance at which the maximum key generation rate for
certain protocol hits zero is defined as maximum secure distance for
this protocol on this set-up. It would probably be a waste of time
to perform certain decoy state protocol far beyond its maximum
secure distance.

We performed numerical simulation based on the set-up on which we
implemented the one-decoy protocol. The result is shown in Fig.
\ref{fig:simulation-one}. The power of decoy method is explicitly
shown by the fact that the maximum distance in absence of decoy
method is only 9.5km. In other words, at 15km, not even a single bit
could be shared between Alice and Bob with guaranteed security. In
contrast, with decoy states, our QKD set-up can be made secure over
60km, which is substantially larger than the secure distance (9.5km)
without decoy states.

\begin{figure}[!t]\center
  \includegraphics[width=7cm]{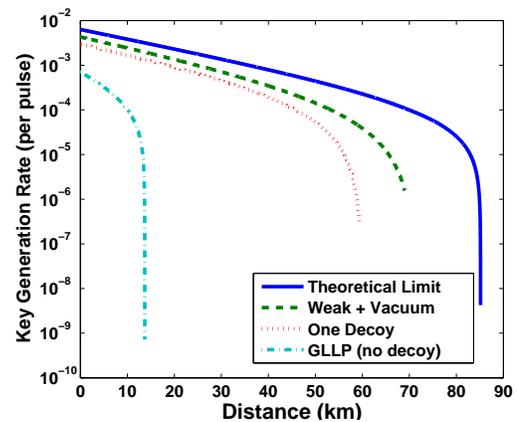}
  \caption{Simulation result of the set-up on which we implemented the
  weak+vacuum
  protocol. The intrinsic parameters of this set-up is shown in the third row of Table
  \ref{intrisic}. Note that this set-up is different from the one we
  implemented one-decoy as reflected by the fact that in Table \ref{intrisic}, the
  values in row 3 are different from the values in row 2.
   Solid line: the theoretical limit of key generation rate. Its maximum transmission
  distance is about 84km. Dashed line: the performance of weak+vacuum protocol. Its maximum distance is about 68km.
  Dotted line: the performance of one-decoy protocol. Its maximum distance is about 59km. Dashed and dotted line:
  the performance without decoy method. Its maximum distance is only 14km.}\label{fig:simulation-wv}
\end{figure}

The set-up on which we implemented the weak+vacuum protocol is a bit
different from the one we implemented the one-decoy protocol because
the single photon detector had been adjusted by the manufacturer and
several important properties, including $\eta_{Bob}$, $Y_0$ and
$e_{detector}$, were changed, as shown in Table \ref{intrisic}. The
simulation result for this ``new'' set-up is shown in Fig.
\ref{fig:simulation-wv}. Clearly, the expected performance,
including the key rate of certain distance and maximum secure
distance of certain protocol, of this set-up is different from the
previous one. This difference is natural because the properties of
the system have changed.

The advantage of weak+vacuum protocol over one-decoy protocol is
shown by the fact that the maximum secure distance of one-decoy
protocol is 59km, which means that one-decoy protocol cannot give
out a positive key rate at 60km. We confirmed this numerical
simulation result by plugging experimental results $Q_\mu$, $E_\mu$,
$Q_\nu$ and $q$ from Table \ref{tab:expresult_wv} into Eqs.
\eqref{eq:Q1e1_onedecoy}\eqref{eq:QnuL}\eqref{eq:RL} and found
indeed that key rate is not positive.

The maximum secure distance of our set-up is limited by equipment,
especially the single photon detectors we used (APDs in Figs.
\ref{schematic_one}\&\ref{schematic_wv}). Given a better set-up
(higher $\eta_{Bob}$, lower $e_{detector}$ and $Y_0$), secure decoy
state QKD can be experimentally implemented over 100km, as shown in
\cite{practical}.

\section{Conclusion}

For the first time, we have implemented decoy state QKD. We have
implemented two protocols: the one-decoy protocol and the
weak+vacuum protocol. Simple modifications (adding AOMs) on a
commercial QKD system are made to implement decoy state QKD. The
simplicity of the modification (much simpler than building a
near-perfect single photon source) shows the feasibility of decoy
method. Also, the high key rates and long transmission distances
(60km) show the power of decoy method. Given better QKD set-ups,
decoy state method could make secure QKD at even longer distances.

We thus come to the conclusion: decoy method is ready for immediate
real-life applications!

\section*{Acknowledgment}
% optional entry into table of contents (if used)
%\addcontentsline{toc}{section}{Acknowledgment}
We thank generous help from many colleagues including Gr\'{e}goire
Ribordy. Support of the funding agencies CFI, CIPI, the CRC program,
NSERC, and OIT is gratefully acknowledged.



% http://www.ctan.org/tex-archive/macros/latex/contrib/supported/IEEEtran/bibtex
%\bibliographystyle{IEEEtran.bst}

\end{document}